\documentclass[11pt,a4paper]{article}

\pdfoutput=1

\usepackage{comment}
\usepackage{jcappub}

\usepackage{physics}
\usepackage{subfig}
\usepackage{float}
\usepackage{xcolor}

\newcommand{\be}{\begin{equation}}
\newcommand{\ee}{\end{equation}}
\newcommand{\bea}{\begin{eqnarray}}

\newcommand{\eea}{\end{eqnarray}}
\newcommand{\nn}{\nonumber}
\newcommand{\pd}{\partial}
\newcommand{\MP}{M_\text{P}}

\newcommand{\FTtwo}{F_{>2}(R)}

\title{\centering Beyond (and back to) Palatini quadratic gravity and inflation}

\author[a,b]{Christian Dioguardi}
\author[b]{Antonio Racioppi}
\author[b]{Eemeli Tomberg}

\affiliation[a]{Tallinn University of Technology, Akadeemia tee 23, 12618 Tallinn, Estonia}
\affiliation[b]{National Institute of Chemical Physics and Biophysics, R\"avala 10, 10143 Tallinn, Estonia}

\emailAdd{christian.dioguardi@kbfi.ee}
\emailAdd{antonio.racioppi@kbfi.ee}
\emailAdd{eemeli.tomberg@kbfi.ee }

\abstract{We study single-field slow-roll inflation embedded in Palatini $F(R)$ gravity where $F(R)$ grows faster than $R^2$. Surprisingly, the consistency of the theory requires the Jordan frame inflaton potential to be unbounded from below. Even more surprisingly, this corresponds to an Einstein frame inflaton potential bounded from below and positive definite. We prove that for all such Palatini $F(R)$'s, there exists a universal strong coupling limit corresponding to a quadratic $F(R)$ with the $\emph{wrong}$ sign for the linear term and a cosmological constant in the Jordan frame. In such a limit, the tensor-to-scalar ratio $r$ does not depend on the original inflaton potential, while the scalar spectral index $n_s$ does. Unfortunately, the system is ill-defined out of the slow-roll regime. A possible way out is to upgrade to a $F(R,X)$ model, with $X$ the Jordan frame inflaton kinetic term. Such a modification essentially leaves the inflationary predictions unaffected.}


\begin{document}
\maketitle

\section{Introduction}
Several observations of the cosmic microwave background radiation (CMB) support the idea of a flat and homogeneous Universe  at large distances. Such features can be explained by assuming the Universe undergoes accelerated expansion during its very early stages \cite{Starobinsky:1980te,Guth:1980zm,Linde:1981mu,Albrecht:1982wi}. This inflationary era can also generate and preserve the primordial inhomogeneities which generated the subsequent large-scale structure that we observe. In its minimal version, the inflaton, a scalar particle embedded in Einsteinian gravity, drives the near-exponential expansion via its quasi-constant potential energy density.

On the other hand, non-minimal models give more freedom in formulating the theory and exploring the available parameters space (e.g. \cite{Jarv:2016sow} and references therein). Among them, models non-minimally coupled to gravity in the Palatini formulation  have received a lot of attention recently, e.g. \cite{Tamanini:2010uq,Bauer:2010jg,Rasanen:2017ivk,Tenkanen:2017jih,Racioppi:2017spw,Markkanen:2017tun,Jarv:2017azx,Racioppi:2018zoy,Kannike:2018zwn,Enckell:2018kkc,Enckell:2018hmo,Rasanen:2018ihz,Bostan:2019uvv,Bostan:2019wsd,Carrilho:2018ffi,Almeida:2018oid,Takahashi:2018brt,Tenkanen:2019jiq,Tenkanen:2019xzn,Tenkanen:2019wsd,Kozak:2018vlp,Antoniadis:2018yfq,Antoniadis:2018ywb,Gialamas:2019nly,Racioppi:2019jsp,Rubio:2019ypq,Edery:2019txq,Lloyd-Stubbs:2020pvx,Das:2020kff,McDonald:2020lpz,Shaposhnikov:2020fdv,Enckell:2020lvn,Jarv:2020qqm,Gialamas:2020snr,Karam:2020rpa,Gialamas:2020vto,Karam:2021wzz,Karam:2021sno,Gialamas:2021enw,Annala:2021zdt,Racioppi:2021ynx,Cheong:2021kyc,Mikura:2021clt,Ito:2021ssc,Racioppi:2021jai,AlHallak:2021hwb,AlHallak:2022gbv,Gialamas:2022gxv,Dimopoulos:2022rdp,Dimopoulos:2022tvn,Gialamas:2022xtt}.
In  the more usual metric formulation, the  metric tensor is the only dynamical degree of freedom, while the connection is chosen to be the Levi-Civita one.  On the contrary,  in the Palatini formulation, both the metric and the connection are dynamical variables. Their corresponding equations of motion (EoMs) will set their eventual relation. In the case of the Einstein--Hilbert action, the theories become equivalent, i.e., the Levi-Civita connection is a consequence of the EoMs. Otherwise, in the presence of non-minimal couplings to gravity, the theories are completely different and lead to different phenomenological predictions, e.g. \cite{Koivisto:2005yc,Bauer:2008zj}.

In this article, we are interested in a particular class of non-minimal Palatini models: the $F(R)$ models. A specific choice,  $F(R)=R+\alpha R^2$, was already studied in \cite{Enckell:2018hmo}, while a more general study was presented in \cite{Dioguardi:2021fmr}. In the previously studied $F(R)=R+\alpha R^2$, the EoM for the auxiliary field (related to the connection) is independent of $\alpha$ and simple to solve \cite{Enckell:2018hmo}. On the contrary, it is not always possible to solve the constraint equation of the auxiliary field analytically for an arbitrary $F(R)$. Therefore,  a new method was introduced in \cite{Dioguardi:2021fmr} that allows to circumvent this issue and still compute the inflationary observables. The only requirement was that the solution to the auxiliary field equation exists in the first place. It was also discovered that if the Jordan frame inflaton potential is positive semidefinite and unbounded from above, the existence of such a solution requires that $F(R)$ cannot diverge faster than $R^2$ at high curvature values.

The purpose of this work is to study the opposite case. We consider a Jordan frame inflaton potential that is unbounded from below. The existence of a solution for the EoM of the auxiliary field allows only $F(R)$ that diverges faster than $R^2$ at high curvature values (from now on, we use the notation $\FTtwo$ for a $F(R)$ satisfying this property). Surprisingly, we will see that the corresponding Einstein frame potential is positive definite and bounded from above, making it a good candidate for inflation \emph{\'a la} hilltop (e.g. \cite{Boubekeur:2005zm,Tzirakis:2007bf,Lillepalu:2022knx} and refs. therein). Even more interestingly, in the strong coupling limit, inflation can be understood in general terms without specifying the particular form of the $F(R)$: in such a limit, under slow-roll, any $\FTtwo$ behaves quadratically as $ F(R)=2\Lambda - \omega R + \alpha R^2$. The unfamiliar negative sign for the Einstein--Hilbert term (still allowed as long as the constraints  $F'(R)>0$, $F''(R) \neq 0$ hold) plays a key role in solving the EoM of the auxiliary field for negative Jordan frame inflaton potentials. Unfortunately, $F(R)$ models alone seem to be unable to provide a graceful exit from the inflationary era. For this reason, we briefly study an extension of these theories, namely $F(R,X)$ where $X$ is the inflaton kinetic term, showing how this class of models can solve the main issues found for the simple $F(R)$ case while keeping the same inflationary predictions in the strong coupling limit.

This article is organized as follows. In Section \ref{sec:beyond:R2}, we introduce the model and discuss the necessity of a negative potential for the consistency of the theory.  In Section \ref{sec:test}, we present some numerical examples for different choices of $\FTtwo$'s and Jordan frame inflaton potentials. Moreover, we also show how any kind of $\FTtwo$ converges to a specific type of quadratic gravity in the strong coupling limit. In Section~\ref{sec:beyondSR}, we study the model's behavior beyond the slow-roll limit and find that a graceful exit does not take place. In Section \ref{sec:beyond_F(R)}, we extend our discussion to $F(R,X)$ models and focus on how they solve the issues of the $F(R)$ theories. Finally, we present our conclusions in Section \ref{sec:Conclusions}.

\section{Palatini gravity and negative potentials} \label{sec:beyond:R2}

\begin{figure}[t]
    \centering
    \includegraphics[width=0.45\textwidth]{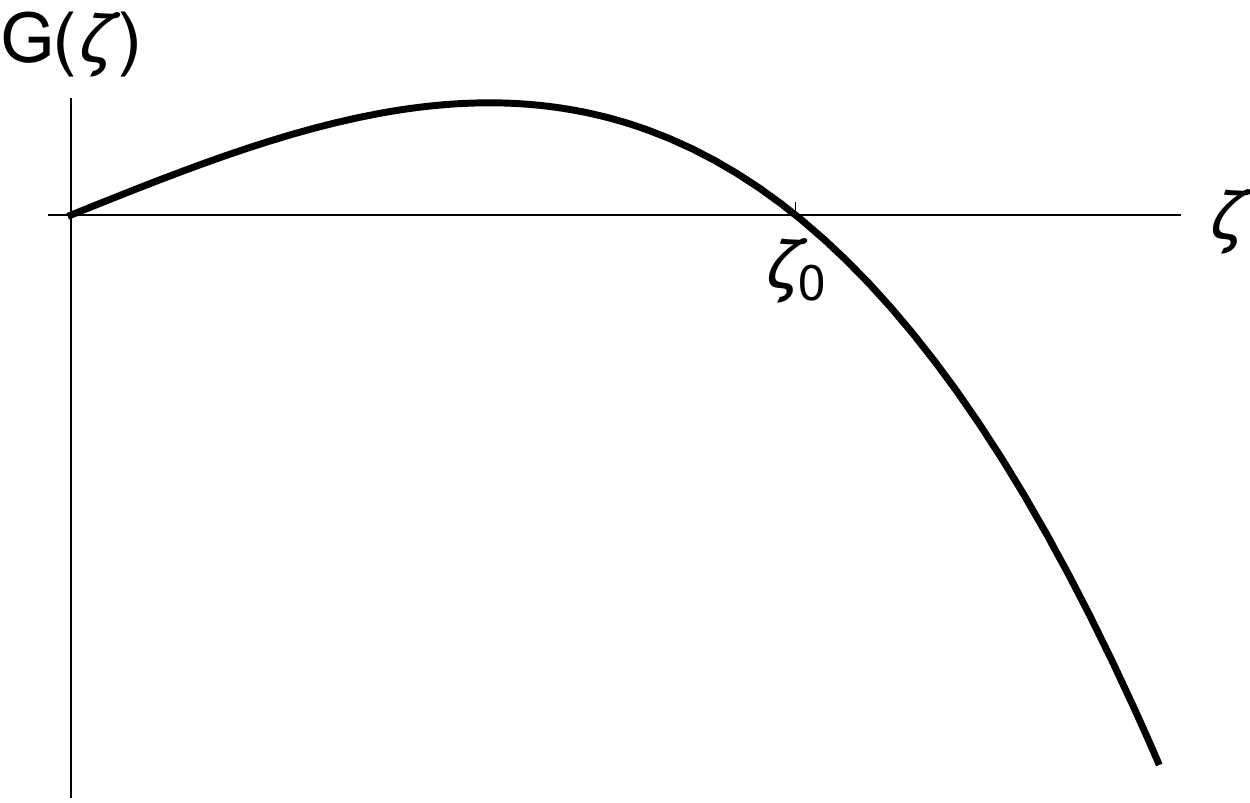}
    \qquad
    \includegraphics[width=0.45\textwidth]{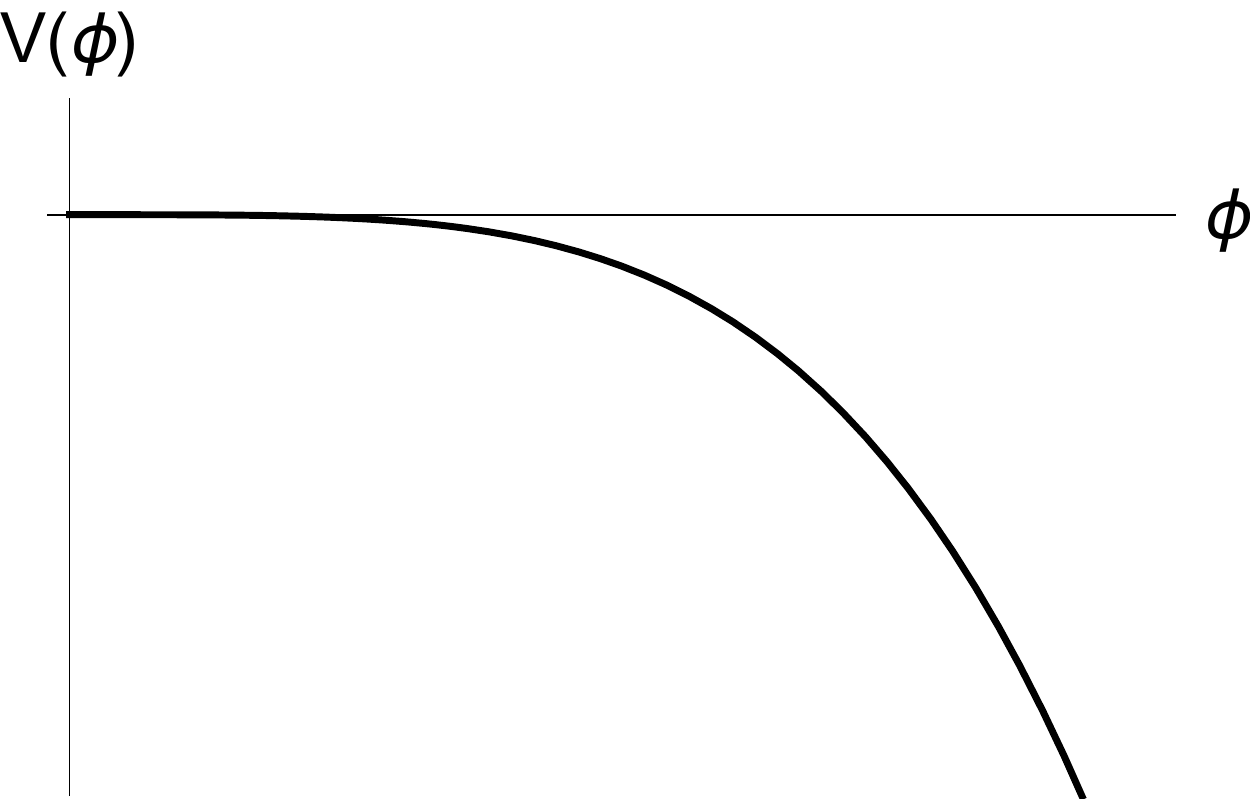}
    \caption{Reference plots of a $G(\zeta)$ (left) generated by a $\FTtwo$ and of a negative $V(\phi)$ (right). Notice that $G(\zeta)=V(\phi)$ is satisfied easily for any $\phi$ when $\zeta \geq \zeta_0$.}%
    \label{fig:GvsV}
\end{figure}

We start with the following action for a real scalar inflaton $\phi$ minimally coupled to $F(R)$ gravity (we assume Planck units, $\MP=1$, and a space-like metric signature):
\be \label{eq:actionFR}
  S_J = \int \dd^4 x \sqrt{-g_J} \left[\frac{1}{2} F(R(\Gamma)) - \frac{1}{2} g_J^{\mu\nu} \partial_\mu \phi \partial_\nu \phi - V(\phi) \right] \, ,
\ee
where the notation $R(\Gamma)$ for the curvature scalar emphasizes that we are considering the Palatini formulation of gravity and $\Gamma^\rho_{\mu\nu}$ is the connection in the Jordan frame. The generic setup has been extensively studied in \cite{Dioguardi:2021fmr}; in the following, we will recap the most important details.
As is customary, we rewrite the action \eqref{eq:actionFR} using an auxiliary field $\zeta$ as
\bea \label{eq:action:zeta:J}
  S_J &=& \int \dd^4 x \sqrt{-g}_J \left[\frac{1}{2} \left[F(\zeta)+F'(\zeta) \left(R(\Gamma)-\zeta \right) \right] - \frac{1}{2}  g_J^{\mu\nu} \partial_\mu \phi \partial_\nu \phi - V(\phi) \right] \, ,
\eea
where we used $F'(\zeta)\equiv\partial F/\partial \zeta$. Then, we move to the Einstein frame via the Weyl transformation
\be \label{eq:Weyl}
  g_{E\, \mu \nu} = F'(\zeta)  \ g_{J \, \mu \nu} \, ,  
\ee
obtaining
\be \label{eq:action:zeta:E}
  S_E = \int \dd^4 x \sqrt{-g_E} \left[ \frac{1}{2} R_E - \frac{1}{2} g_E^{\mu\nu} \partial_\mu \chi \partial_\nu \chi - U(\chi,\zeta) \right] \, ,
\ee
where the canonically normalized scalar $\chi$ and the scalar potential are, respectively, \cite{Dioguardi:2021fmr}
\bea
\label{eq:dchidphi}
 \frac{\pd \chi}{\pd \phi} &=& \sqrt{\frac{1}{F'(\zeta)}} \, ,\\
 U(\chi,\zeta) &=& \frac{V(\phi(\chi))}{F'(\zeta)^2} - \frac{F(\zeta)}{2 F'(\zeta)^2} + \frac{\zeta}{2 F'(\zeta)} \, .
 \label{eq:Uchizeta}
\eea
By varying \eqref{eq:action:zeta:E} with respect to $\zeta$, we get its EoM in the Einstein frame,
\be
G(\zeta) - \frac{1}{4} F'(\zeta)\partial_\mu \phi\partial^\mu \phi = V(\phi) \, , \label{eq:EoMzetafull}
\ee 
where
\be \label{eq:G}
G(\zeta) \equiv \frac{1}{4} \left[ 2 F(\zeta) - \zeta F'(\zeta) \right] \, .
\ee
Applying the computational strategy introduced in \cite{Dioguardi:2021fmr}, one can show that under slow-roll, the Einstein frame inflaton potential can be formally written as a function of $\zeta$ only (where $\zeta$ is a function of $\chi$ via \eqref{eq:dchidphi}  and the slow-roll version of \eqref{eq:EoMzetafull}):
\be \label{eq:U}
 U(\zeta) =\frac{1}{4} \frac{\zeta}{F'(\zeta)} \, .
\ee
Consistency of the theory requires $F'(\zeta)>0$. If this constraint and the slow-roll version of \eqref{eq:EoMzetafull} are both satisfied and $\zeta$ is positive, then $U$ is positive definite for any $V(\phi)$, positive or negative. The case of a positive $V(\phi)$ has already been studied in \cite{Dioguardi:2021fmr} leading to the conclusion that a theory with an unbounded from above $V(\phi)$ is consistent only if $F(R)$ does not diverge faster than $R^2$ at high curvature values. In this article, we study the opposite configuration and assume a $\FTtwo$. We prove now that a negative $V(\phi)$ is a sufficient condition for the consistency of the theory. First of all, we recall the EoM for $\zeta$ under slow-roll \cite{Dioguardi:2021fmr}:
\be \label{eq:EoMzeta:SR}
G(\zeta) =  V(\phi)  \, ,
\ee
where $G(\zeta)$ is given by \eqref{eq:G}.
Assuming a $\FTtwo$ we can easily check that there exists a certain value $\zeta_0$ so that $G(\zeta_0)=0$ and for $\zeta$ larger (smaller) than $\zeta_0$, we have $G(\zeta)$ negative (positive) and $G(+\infty) \to -\infty$. Therefore, the simplest configuration that keeps the theory consistent is having a negative\footnote{The other possible (but more tuned) scenario where $V(\phi)$ has a local maximum lower than the local maximum of $G(\zeta)$ will be studied in a separate work \cite{toappear.}} $V(\phi)$ (see also Fig.~\ref{fig:GvsV}). Note that such a configuration automatically implies that $\zeta \geq \zeta_0$. In the following subsections, we will study inflation in different realizations of such a scenario.

\begin{figure}[t]
    \centering
    \includegraphics[width=0.45\textwidth]{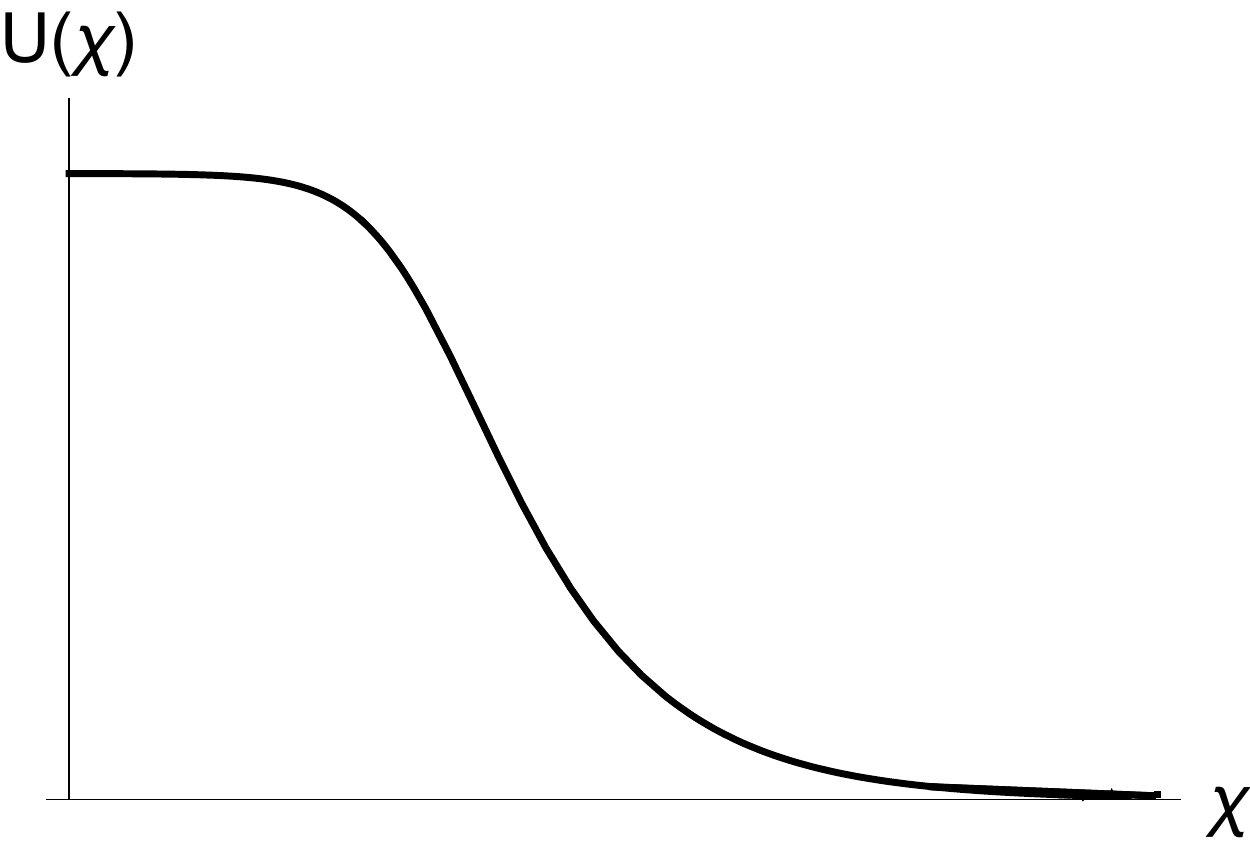}
    \caption{Reference plot for $U(\chi)$ as a function of $\chi$.}
    \label{fig.plot.Vbar}
\end{figure}

\section{Test scenarios} \label{sec:test}
We study the phenomenology of the two simple test cases of a negative monomial potential
\be
V(\phi)= V_k(\phi)=- \lambda_k \phi^k \, , \qquad \lambda_k=\frac{\lambda^k}{k!}\, , \label{eq:V:k} 
\ee
and a negative exponential potential
\be
V(\phi)= V_e(\phi)=- e^{\lambda\phi}= - \sum_{k=0}^{\infty}\frac{\lambda^k}{k!} \phi^k = - \sum_{k=0}^{\infty} \lambda_k\phi^k \, . \label{eq:V:exp}
\ee
The unconventional normalization of \eqref{eq:V:k} is needed in order to make a better comparison between the results of $V_k$ and $V_e$.

We consider three types of $\FTtwo$'s. In order to keep the discussion easy to follow, we will perform numerical analyses only on $F(R)$'s with one free parameter and the prefactor of the Einstein--Hilbert term equal to one. Our choices are: $F(R) = R + \alpha R^n$ (with numerical analysis for $n=3$), $F(R) = \frac{1}{\alpha}e^{\alpha R}$, and $F(R) = R + \alpha R^2 \ln({\alpha R})$. 
For all the cases, the generic behavior of $U$ can be easily derived and is depicted in Fig.~\ref{fig.plot.Vbar}. Since the field redefinition from $\phi$ to $\chi$ is monotonic (see eq. \eqref{eq:dchidphi}), the generic behaviour of $U$ as a function $\phi$ and as a function of $\chi$ are the same. At small $\phi$ ($\chi$) values, when $\zeta \approx \zeta_0$, $U$ has a local maximum, then it decreases for an increasing $\zeta$ with a horizontal asymptote\footnote{We also notice that such a feature gives the possibility of a common explanation of inflation and the late-time value of the cosmological constant. However, such a study is beyond the scope of the current paper, and we postpone it to future work.} at $U=0$.

A detailed analysis of each case for $N_e=60$, where $N_e$ is the number of e-folds between the CMB scale and the end of inflation, is presented in the following\footnote{The case of $N_e=50$ is not discussed here because, as we will see later in Section \ref{sec:back:to:R2}, such a configuration is strongly disfavoured.}. In all cases, $\phi$ ($\chi$)  and $\zeta$ slowly roll towards larger values, down the slope of $U$, as inflation proceeds in the $\zeta>\zeta_0$ regime.

\subsection{$F(R) = R + \alpha R^n$}

Following the procedure of \cite{Dioguardi:2021fmr}, we give the analytic expression for inflationary observables as a function of $\zeta_N$ for $V(\phi)=V_k(\phi)$:
\begin{align} 
& N_e = \int_{\zeta_f}^{\zeta_N} {\rm d}\zeta \frac{16^{-1/k} \lambda _k^{-2/k} \left(\zeta _N-\alpha  (n-2) n \zeta _N^n\right) \left(\alpha 
   (n-2) \zeta _N^n-\zeta _N\right)^{\frac{2}{k}-2}}{k^2} \, ,\\
& \quad\nn\\
& r = \frac{2^{\frac{4}{k}+3} k^2 \lambda _k^{2/k} \left(\alpha  (n-2) \zeta _N^n-\zeta
   _N\right)^{2-\frac{2}{k}}}{\zeta _N \left(\alpha  n \zeta _N^n+\zeta _N\right)}\, , 
\end{align}
\begin{align}
& n_s = 1+\frac{16^{\frac{1}{k}} k \lambda _k^{2/k} \left(\alpha  (n-2) \zeta _N^n-\zeta
   _N\right)^{\frac{k-2}{k}} \left(\alpha  (n-2) (2 (k-1) n-3 k) \zeta _N^n+(k+2) \zeta
   _N\right)}{\zeta _N \left(\zeta _N-\alpha  (n-2) n \zeta _N^n\right)} \, , \\
& A_s = \frac{4^{-\frac{2 (k+1)}{k}} \lambda _k^{-2/k} \zeta _N^3 \left(\alpha  (n-2) \zeta _N^n-\zeta
   _N\right)^{\frac{2}{k}-2}}{3 \pi ^2 k^2} \, .
\end{align}

Unfortunately, not much information can be extracted from the exact results. To simplify the expressions, we give below the analytical formulas for the CMB observables in the strong and weak coupling regimes. Then we show the full numerical predictions for different choices of the Jordan frame potential $V(\phi)$.

Let us start with the large field limit  $\zeta_N \gg \zeta_0$ (the \emph{small} $\alpha$ limit, as we will see later). For the current model, we have
\be
\zeta_0 = \qty(\alpha(n-2))^{\frac{1}{1-n}} \, . \label{eq:z0:Rn}
\ee
In this limit, we can neglect the linear term in $F(R)$ and \eqref{eq:EoMzeta:SR} becomes
\begin{equation}
    \frac{\alpha}{4}(2-n)\zeta^n = -\lambda_k\phi^k \, .
\end{equation}
The analytic formulas for the CMB observables become:
\begin{eqnarray}
 N_e &&= \frac{n \zeta_0^{\frac{2+A}{k}}}{2^{\frac{4}{k}}\lambda^{\frac{2}{k}} k A} \zeta_N^{-\frac{A}{k}} \, , \label{eq:Ne:Rn:infinity}
\\
r &&= \frac{8 k (n-2)}{A}\frac{1}{N_e} \, , \label{eq:r:Rn:large:zeta}
\\
n_s &&= 1 - \frac{B}{A}\frac{1}{N_e} \, , \label{eq:ns:Rn:large:zeta}
\\
A_s &&= \frac{(4\lambda)^\frac{2(n-2)}{A}\zeta_0^\frac{(k-4)(n-1)}{A}}{48\pi^2 k^2}\qty(\frac{k A N_e}{n})^\frac{B}{A} \, ,
\end{eqnarray}
where $A= nk - 2n - k$, $B= 2nk - 2n - 3k$, and we assumed $k\geq 4$ and $n>2$. 

For smaller values of $k$, we have a change in the primitive function coming from the computation of $N_e$. However, as we will see shortly, only results with $k > 4$ may be within the allowed constraints \cite{BICEP:2021xfz}. Therefore, we won't study the $k<4$ case.

\begin{figure}[t]
    \centering
    
    \subfloat[]{\includegraphics[width=0.45\textwidth]{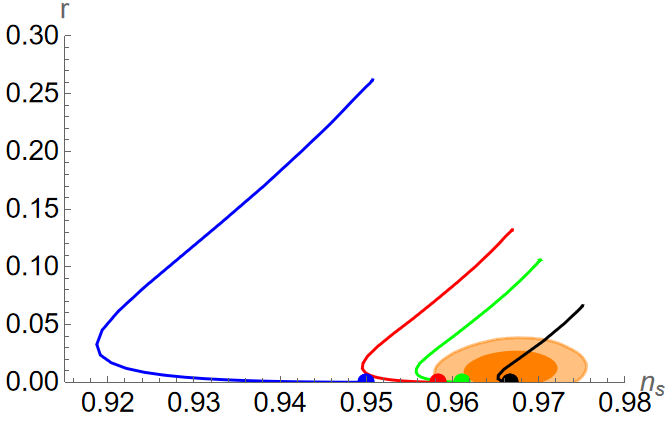}}%
    \qquad
    \subfloat[]{\includegraphics[width=0.45\textwidth]{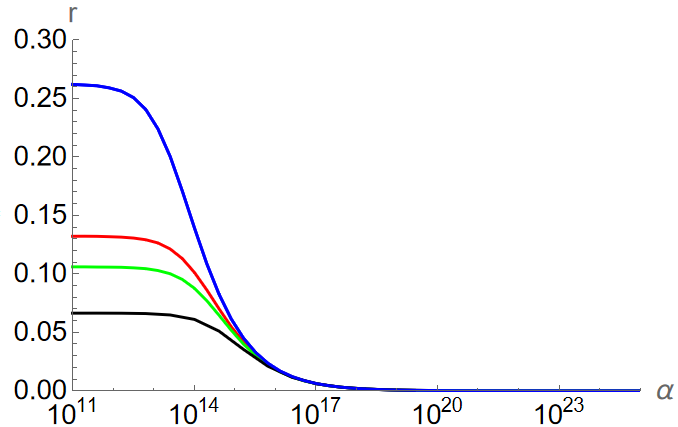}}%
    
    \subfloat[]{\includegraphics[width=0.45\textwidth]{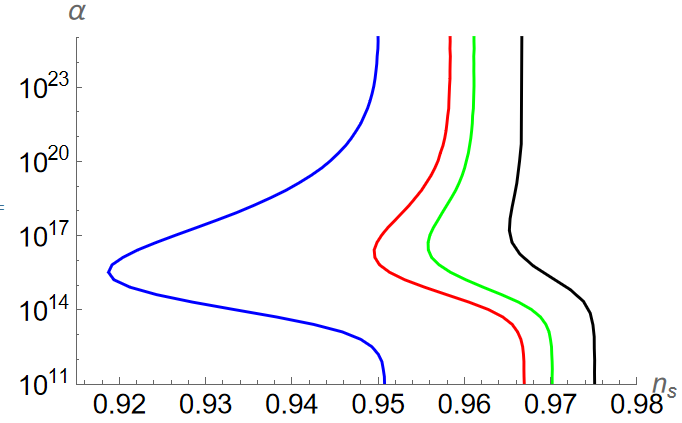}}%
    \qquad
    \subfloat[]{\includegraphics[width=0.45\textwidth]{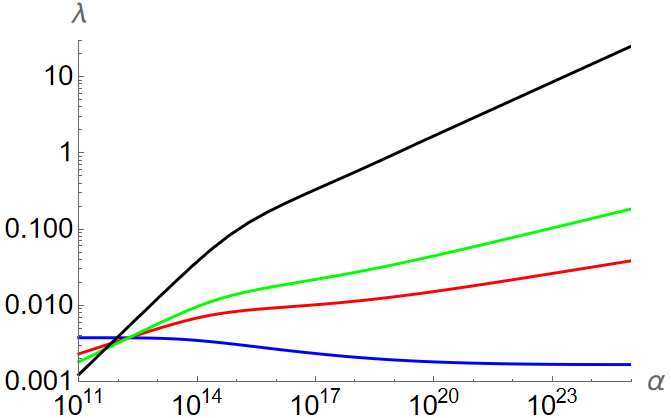}}

    \caption{The observables $r$ vs. $n_s$ (a), $r$ vs. $\alpha$ (b), $\alpha$ vs. $n_s$  (c) and $\lambda$ vs. $\alpha$ (d) for $n=3$ and $V_k$ with $k=4$ (blue), $k=6$ (red), $k=8$ (green) and $V_e$ (black) for $N_e=60$. In the same color code, we show the limit values for $\alpha \rightarrow \infty$ represented by bullets. The scalar amplitude $A_s$ is fixed to its observed value \cite{Planck2018:inflation}.
    The orange areas represent the 1,2$\sigma$ allowed regions coming  from  the latest combination of Planck, BICEP/Keck and BAO data \cite{BICEP:2021xfz}}
    \label{fig.r_vs_ns}
    
\end{figure}
We now move on to the $\zeta\rightarrow \zeta_0$ limit. The analytic results are 
\begin{eqnarray}
r &&= \frac{4k(n-2)}{k-2}\frac{1}{N_e}\qty(\frac{\zeta_N}{\zeta_0}-1)\sim 0 \, , \label{eq:r:Rn:small:zeta}
\\
n_s &&= 1 - \frac{
k-1}{k-2}\frac{2}{N_e} \, , \label{eq:ns:Rn:small:zeta}
\\
A_s &&= \frac{16 (k-2) N_e}{3\pi^2 k (n-1)}\frac{\zeta_0^2}{\zeta_N-\zeta_0} \, .
\label{eq:As:Rn:small:zeta}
\end{eqnarray}
We notice that $A_s$ diverges as $\zeta_N \rightarrow \zeta_0$; hence we must require that $\zeta_0\rightarrow 0$ in order to achieve a finite $A_s \simeq 2.1\cdot 10^{-9}$ \cite{Planck2018:inflation} in the $\zeta\rightarrow \zeta_0$ limit. Given \eqref{eq:z0:Rn}, the limit $\zeta_N \rightarrow \zeta_0$ is then equivalent to the limit $\alpha\rightarrow \infty$.  
It is also important to ensure that the Einstein frame potential energy density does not reach values above the Planck scale, at least during the inflationary era. In the Einstein frame, the maximum potential energy density, i.e the plateau value $U(\zeta_0)$,
\begin{equation}
   U(\zeta_0)= \frac{\zeta_0}{4 F'(\zeta_0)} \simeq \frac{n-2}{8(n-1)} \Big[\alpha(n-2) \Big]^\frac{1}{1-n} \, ,
\end{equation}
actually decreases when $\alpha$ increases (and the necessary condition $n>2$ is satisfied, see discussion around eq. \eqref{eq:EoMzeta:SR}). Therefore the requirement of a sub-Planckian Einstein frame potential energy imposes only the lower bound
\begin{equation}
   \alpha > \frac{1}{n-2} \left[\frac{n-2}{8 (n-1)}\right]^{n-1}\, , \label{eq:lower:alpha:n}
\end{equation}
which is always satisfied when the constraint $A_s \simeq 2.1\cdot 10^{-9}$ \cite{Planck2018:inflation} holds.

Notice as well that $n_s$ is independent of the choice of the exponent $n$ and only depends on the exponent of the monomial potential $k$, while the scalar-to-tensor ratio $r$ is suppressed. 

The results for $V_e(\phi)$ can be derived similarly and, when it comes to $r$ and $n_s$, they are equivalent to the corresponding $k \to \infty$ limit of the $V_k$ results.

We show in Fig.~\ref{fig.r_vs_ns} the numerical results for the observables $r$ vs. $n_s$ (a), $r$ vs. $\alpha$ (b), $\alpha$ vs. $n_s$ (c) and $\lambda$ vs. $\alpha$ (d) for $n=3$ and $V_k$ with $k=4$ (blue), $k=6$ (red), $k=8$ (green) and $V_e$ (black) for $N_e=60$. In the same color code, we show the limit values for $\alpha \rightarrow \infty$ represented by bullets. The scalar amplitude $A_s$ is fixed to its observed value \cite{Planck2018:inflation}.
The orange areas represent the 1,2$\sigma$ allowed regions coming  from  the latest combination of Planck, BICEP/Keck and BAO data \cite{BICEP:2021xfz}.
From (a) we notice that the $V_8$ and $V_e$ potentials predict $r, n_s$ inside the $2\sigma$ region with $r\sim 0$, $n_s \sim 0.961 $ and $n_s \sim 0.967$ (in the strong coupling limit). From (b), we notice that all potentials have a suppressed $r$ in this limit, as expected from the analytical computations. The suppression starts to become effective around $\alpha\sim 10^{14}$, and for $\alpha \gtrsim 10^{17}$, all potentials predict essentially the same $r$. From (c), we see that all potentials exhibit similar behavior in $n_s$ which decreases to a minimum value around $\alpha \sim 10^{16}$ and then progressively increases towards the asymptotic value \eqref{eq:ns:Rn:small:zeta}. 
Finally, from (d) we have the relation between the two model parameters $\lambda,\alpha$ obtained by fixing $A_s$ to the observed value. We notice that for all potentials, increasing $\alpha$ implies increasing $\lambda$, with the exception of the $V_4$ potential for which $\lambda$ decreases. 

\subsection{$F(R) = \frac{1}{\alpha}e^{\alpha R}$} \label{sec:exp}
The exact solutions for this model are
\bea
&&\hspace{-1cm} N_e =  \int_{\zeta_f}^{\zeta_N} {\rm d}\zeta \frac{2^{-4/k} \lambda _k^{-2/k} }{\zeta _0
   k^2} \left(\zeta _0-2 \zeta _N\right) \zeta _N \left(e^{\frac{2 \zeta
   _N}{\zeta _0}}\right)^{\frac{2}{k}-1} \left(\zeta _N-\zeta _0\right)^{\frac{2}{k}-2} \, ,\\
&&\hspace{-1cm} r = \frac{2^{\frac{4}{k}+3} k^2 \lambda _k^{2/k} \left(e^{\frac{2 \zeta _N}{\zeta _0}} \left(\zeta
   _N-\zeta _0\right)\right)^{1-\frac{2}{k}}}{\zeta _N^2} \, , \label{eq:r:expR}\\
&&\hspace{-1cm} n_s = 1+\frac{16^{\frac{1}{k}} k \lambda _k \left(3 \zeta _0^2 k+\zeta _0 (2-5 k) \zeta _N+4 (k-1) \zeta
   _N^2\right) }{\left(\zeta _0-2 \zeta _N\right) \zeta _N^2} \left(\frac{e^{\frac{2 \zeta _N}{\zeta _0}} \left(\zeta _N-\zeta _0\right)}{\lambda
   _k}\right)^{1-\frac{2}{k}} \, , \\
&&\hspace{-1cm} A_s = \frac{2^{-\frac{4 (k+1)}{k}} \zeta _N^3 }{3 \pi ^2 k^2 \lambda _k^2} \left(\frac{e^{\frac{2 \zeta _N}{\zeta _0}} \left(\zeta
   _N-\zeta _0\right)}{\lambda _k}\right)^{\frac{2}{k}-2} \, ,
\eea 
where we used $\zeta_0=\frac{2}{\alpha}$. From  \eqref{eq:r:expR}, we see that when $\zeta_N \gg \zeta_0$, then the exponential term dominates the equation for $r$, leading to a value not compatible with slow-roll. Therefore the limit $\zeta \gg \zeta_0$ is actually never realized in such a scenario and only the $\zeta\rightarrow\zeta_0$ limit remains:
\bea
r &&= \frac{8 k}{k-2}\frac{1}{N_e}\qty(\frac{\zeta_N}{\zeta_0} - 1) \sim 0 \label{eq:r:eR:small:zeta} \, ,
\\%
n_s &&= 1 - \frac{k-1}{k-2}\frac{2}{N_e} \, , \label{eq:ns:eR:small:zeta}
\\%
A_s &&= \frac{16 (k-2)}{3 k \pi^2 e^2} \frac{\zeta_0^2}{\zeta_N - \zeta_0} \, ,
\label{eq:As:eR:small:zeta}
\eea
where we notice that the equation for $n_s$ is exactly the same as \eqref{eq:ns:Rn:small:zeta}. Moreover, similarly to the previous case, the equation for $A_s$ provides the equivalence between the limits $\zeta_N\rightarrow\zeta_0$ and $\alpha\to\infty$.
The requirement of a sub-Planckian Einstein frame potential $U$ imposes now the lower bound
\begin{equation}
   \alpha > \frac{1}{2 e^2} \, , \label{eq:lower:alpha:exp}
\end{equation}
which is always satisfied when the constraint $A_s \simeq 2.1\cdot 10^{-9}$ \cite{Planck2018:inflation} holds.

The results for $V_e(\phi)$ can be derived in a similar way and, when it comes to $r$ and $n_s$, they are equivalent to the corresponding $k \to \infty$ limit of the $V_k$ results.

\begin{figure}[t]
    \centering
    
    \subfloat[]{\includegraphics[width=0.45\textwidth]{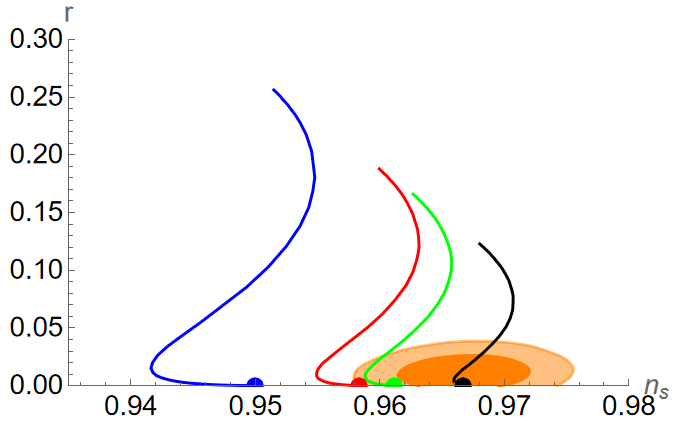}}%
    \qquad
    \subfloat[]{\includegraphics[width=0.45\textwidth]{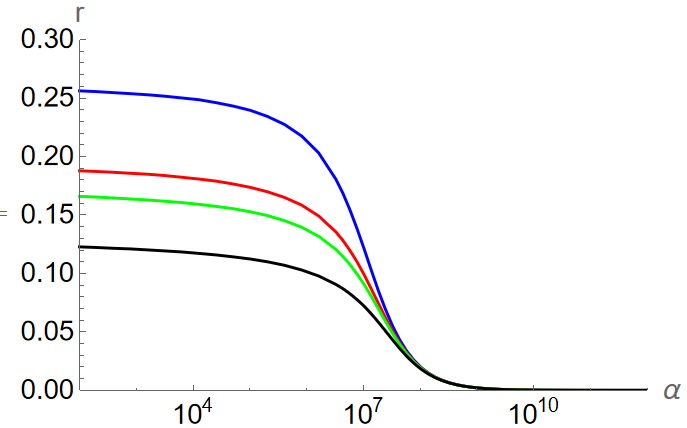}}%
    
    \subfloat[]{\includegraphics[width=0.45\textwidth]{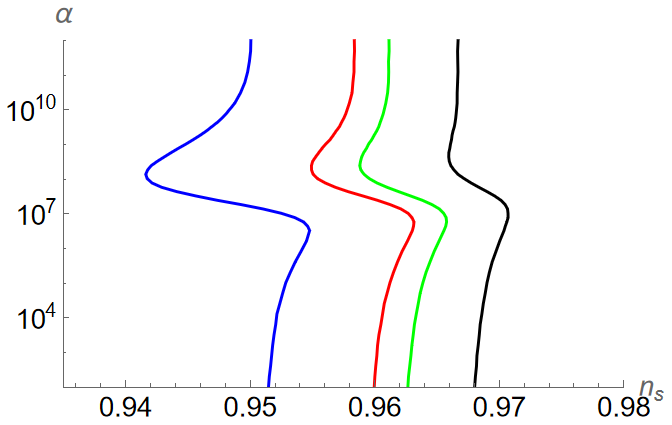}}%
    \qquad
    \subfloat[]{\includegraphics[width=0.45\textwidth]{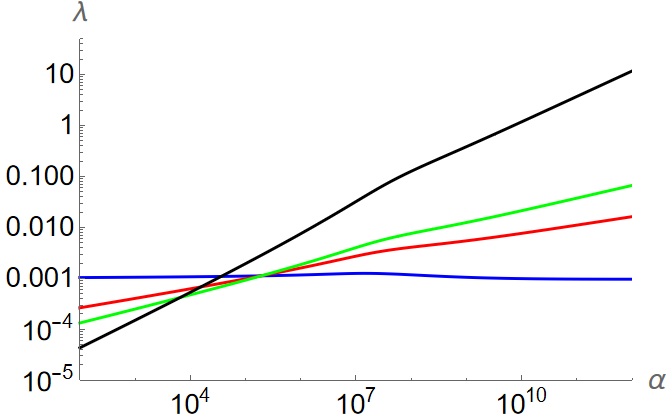}}

    \caption{
    The observables $r$ vs. $n_s$ (a), $r$ vs. $\alpha$ (b), $\alpha$ vs. $n_s$  (c) and $\lambda$ vs. $\alpha$ (d) for $F(R) = \frac{1}{\alpha}e^{\alpha R}$, and $V_k$ with $k=4$ (blue), $k=6$ (red), $k=8$ (green) and $V_e$ (black) for $N_e=60$.  In the same color code, we show the limit values for $\alpha \rightarrow \infty$ represented by bullets.  The scalar amplitude $A_s$ is fixed to its observed value \cite{Planck2018:inflation}.
    The orange areas represent the 1,2$\sigma$ allowed regions coming  from  the latest combination of Planck, BICEP/Keck and BAO data \cite{BICEP:2021xfz}. }
    \label{fig.r_vs_ns_exp}

   \end{figure} 

We show in Fig.~\ref{fig.r_vs_ns_exp} the numerical prediction for the observables $r$ vs. $n_s$ (a), $r$ vs. $\alpha$ (b), $\alpha$ vs. $n_s$  (c) and $\lambda$ vs. $\alpha$ (d) for $F(R) = \frac{1}{\alpha}e^{\alpha R}$, and $V_k$ with $k=4$ (blue), $k=6$ (red), $k=8$ (green) and $V_e$ (black) for $N_e=60$. In the same color code, we show the limit values for $\alpha \rightarrow \infty$ represented by bullets.  The scalar amplitude $A_s$ is fixed to its observed value \cite{Planck2018:inflation}. The orange areas represent the 1,2$\sigma$ allowed regions coming  from  the latest combination of Planck, BICEP/Keck and BAO data \cite{BICEP:2021xfz}. 
From (a) we notice that the $V_8$ and $V_e$ potentials once again predict $r, n_s$ inside the $2\sigma$ region with $r\sim 0$, $n_s \sim 0.961 $ and $n_s \sim 0.967 $ respectively (in the strong coupling limit). 
From (b), we notice that all potentials have a suppressed $r$ in this limit, as expected from the analytical computations. The suppression starts to become effective around $\alpha\sim 10^{6}$, and for $\alpha \gtrsim 10^{8}$, all potentials predict the same $r$. From (c) we see that all potentials exhibit similar behavior in $n_s$. Its value increases up to a maximum around $\alpha \sim 10^7$, then decreases to a minimum around $\alpha \sim 10^{9}$ and finally increases again towards the asymptotic value \eqref{eq:ns:eR:small:zeta}. Notice that the asymptotic value for $n_s$ predicted by $V_e$ is $n_s = 1 - \frac{2}{N_e}$ which corresponds to the $k\rightarrow \infty$ in \eqref{eq:ns:eR:small:zeta}. Finally, from (d) we have the relation between the two model parameters $\lambda,\alpha$ obtained by fixing $A_s$ to the observed value. We notice that for all potentials, increasing $\alpha$ implies increasing $\lambda$ with the exception of the $V_4$ potential for which $\lambda$ decreases.

\subsection{$F(R) = R + \alpha R^2 \ln({\alpha R})$ }

The consistency conditions for this model require $F'(\zeta)>0$ for $\zeta>\zeta_0 = \frac{1}{\alpha}$. This sets the constraint\footnote{Considering a more generic $F(R) = R + \alpha R^2 \ln({\beta R})$ would have set the condition $\beta > \frac{\alpha}{{e}}$.} $\alpha > \frac{\alpha}{{e}}$,  which is always satisfied for any positive $\alpha$. 
The exact solutions for the inflationary observables are
\bea
&&\hspace{-1cm} N_e =  \int_{\zeta_f}^{\zeta_N} {\rm d}\zeta \,  \frac{16^{-1/k} \lambda _k^{-2/k} \left(1-\frac{2 \zeta _N}{\zeta _0}\right) \zeta _N^{\frac{2}{k}-1} \left(\frac{\zeta
   _N}{\zeta _0}-1\right)^{\frac{2}{k}-2}}{k^2}\, ,\\
&&\hspace{-1cm} r = \frac{2^{\frac{4}{k}+3} k^2 \lambda _k^{2/k} \zeta _N^{-2/k} \left(\frac{\zeta _N}{\zeta
   _0}-1\right)^{2-\frac{2}{k}}}{1+\frac{\zeta _N}{\zeta _0}+\frac{2 \zeta _N \ln \left(\frac{\zeta _N}{\zeta
   _0}\right)}{\zeta _0}} \, , \\
&&\hspace{-1cm} n_s = 1-\frac{16^{\frac{1}{k}} k \lambda _k^{2/k} \zeta _N^{-2/k} \left(\frac{\zeta _N}{\zeta _0}-1\right)^{1-\frac{2}{k}}
   \left(\frac{(k-4) \zeta _N}{\zeta _0}+k+2\right)}{\frac{2 \zeta _N}{\zeta _0}-1} \, , \\
&&\hspace{-1cm} A_s = \frac{2^{-\frac{4 (k+1)}{k}} \lambda _k^{-2/k} \zeta _N^{\frac{2}{k}+1} \left(\frac{\zeta _N}{\zeta
   _0}-1\right)^{\frac{2}{k}-2}}{3 \pi ^2 k^2} \, .
\eea 
First of all, we derive the analytical results for $\zeta\gg\zeta_0$. In this limit, for $k>4$, we get:
\bea
N_e &=& \frac{2^{1-\frac{4}{k}}}{\alpha k (k-4)}\qty(\frac{\alpha}{\lambda})^{\frac{2}{k}}\zeta^{\frac{4}{k}-1} \, ,
 \label{eq:As:large:zeta:lnR}
 \\
 r &\ll& 1 \, ,\\ 
 n_s &=& 1 - \frac{1}{N_e} \, ,\\
 A_s &=& \frac{k-4}{96\pi^2 k} \frac{N_e}{\alpha} \, .
\eea
For $k=4$, we have a change of primitive, and once again, the limit $\zeta_N \gg \zeta_0$ cannot be achieved. In fact, this limit leads to $\zeta_N \simeq e^{-\frac{N_e}{96 \pi ^2 \alpha  A_s}}$, which is never much bigger than $\zeta_0$ for any $\alpha$ if $A_s$ and $N_e$ satisfy the experimental constraints \cite{Planck2018:inflation}. It can be shown that for any $\alpha$, the experimental constraints force $\zeta_N \approx \zeta_0$ for $k=4$.

We then consider the case $\zeta\rightarrow\zeta_0$, giving
\bea
r(\zeta) &=& \frac{4k}{(k-2)}\frac{1}{N_e}\qty(\frac{\zeta_N}{\zeta_0}-1) \sim 0 \label{eq:r:R2lnR:small:zeta} \, ,
\\%
n_s(\zeta) &=& 1-\frac{k-1}{k-2}\frac{2}{N_e} \, , \label{eq:ns:R2lnR:small:zeta}
\\%
A_s(\zeta) &=& \frac{16(k-2)N_e}{3\pi^2 k}\frac{\zeta_0^2}{\zeta_N-\zeta_0} \, ,
\label{eq:As:R2lnR:small:zeta}
\eea
where once more the equation for $n_s$ remained exactly the same as \eqref{eq:ns:Rn:small:zeta} and \eqref{eq:ns:eR:small:zeta}.
Similarly to the previous cases, from the equation of $A_s$ we deduce the equivalence between the limits $\zeta_N\rightarrow\zeta_0$ and $\alpha\to\infty$.
The requirement of a sub-Planckian Einstein frame potential $U$ imposes now the lower bound
\begin{equation}
    \alpha > \frac{1}{8} \, , \label{eq:lower:alpha:ln}
\end{equation}
which is always satisfied when the constraint $A_s \simeq 2.1 \cdot 10^{-9}$ \cite{Planck2018:inflation} holds.

The results for $V_e(\phi)$ can be derived similarly and, when it comes to $r$ and $n_s$, they are equivalent to the corresponding $k \to \infty$ limit of the $V_k$ results.

\begin{figure}[t]
    \centering
    
    \subfloat[]{\includegraphics[width=0.45\textwidth]{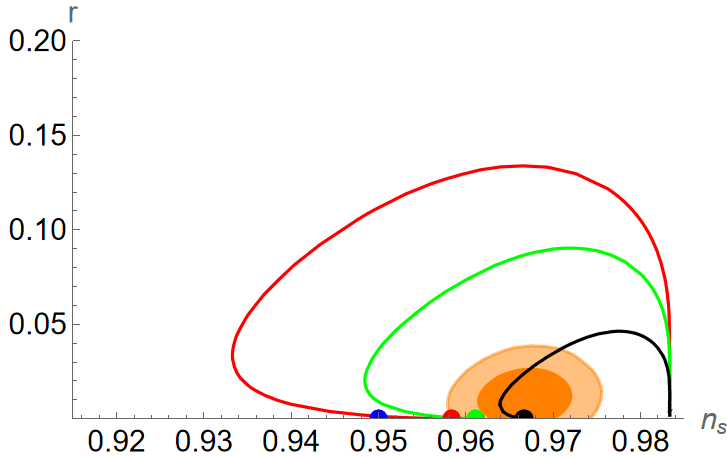}}%
    \qquad
    \subfloat[]{\includegraphics[width=0.45\textwidth]{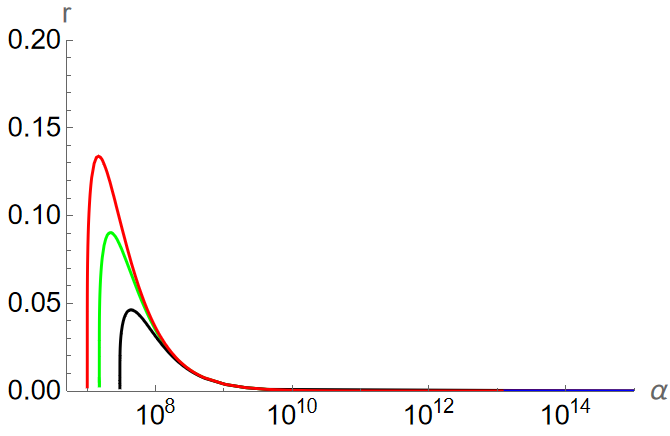}}%
    
    \subfloat[]{\includegraphics[width=0.45\textwidth]{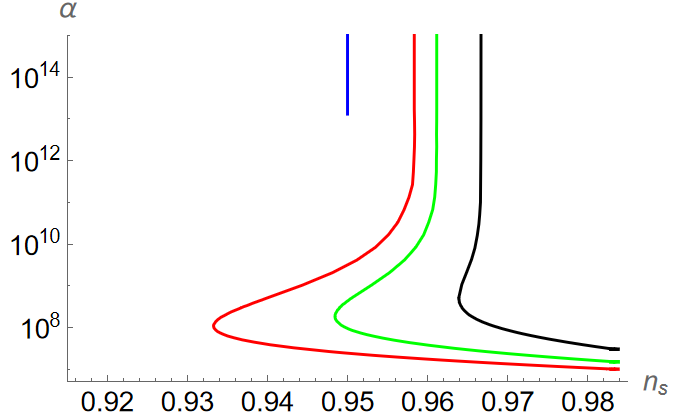}}%
    \qquad
    \subfloat[]{\includegraphics[width=0.45\textwidth]{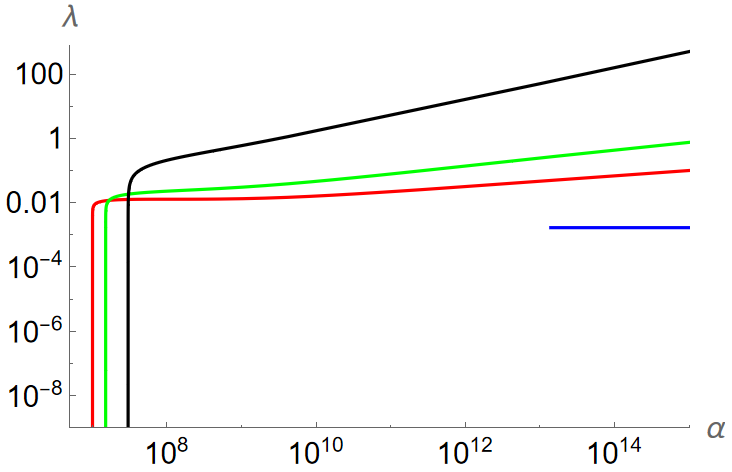}}

    \caption{
    The observables $r$ vs. $n_s$ (a), $r$ vs. $\alpha$ (b), $\alpha$ vs. $n_s$  (c) and $\lambda$ vs. $\alpha$ (d) for $F(R) = R + \alpha R^2 \ln(\alpha R)$, and $V_k$ for $k=4$ (blue), $k=6$ (red), $k=8$ (green) and $V_e$ (black) for $N_e=60$.  In the same color code, we show the limit values for $\alpha \rightarrow \infty$ represented by bullets. The scalar amplitude $A_s$ is fixed to its observed value \cite{Planck2018:inflation}.
    The orange areas represent the 1,2$\sigma$ allowed regions coming  from  the latest combination of Planck, BICEP/Keck and BAO data \cite{BICEP:2021xfz}. }
    \label{fig.r_vs_ns_log}

   \end{figure} 

We show in Fig.~\ref{fig.r_vs_ns_log} the numerical results for the observables $r$ vs. $n_s$ (a), $r$ vs. $\alpha$ (b), $\alpha$ vs. $n_s$  (c) and $\lambda$ vs. $\alpha$ (d) for $F(R) = R + \alpha R^2 \ln(\alpha R)$, and $V_k$ for $k=4$ (blue), $k=6$ (red), $k=8$ (green) and $V_e$ (black) for $N_e=60$. In the same color code, we show the limit values for $\alpha \rightarrow \infty$ represented by bullets.  The scalar amplitude $A_s$ is fixed to its observed value \cite{Planck2018:inflation}. The orange areas represent the 1,2$\sigma$ allowed regions coming  from  the latest combination of Planck, BICEP/Keck and BAO data \cite{BICEP:2021xfz}.
From (a) we notice that the $V_8$ and $V_e$ potentials once more predict $r, n_s$ inside the $2\sigma$ region with $r\sim 0$, $n_s \sim 0.961 $ and $n_s \sim 0.967 $ respectively (in the strong coupling limit). We also notice that the opposite limit gives a suppressed $r$ around $n_s = 0.983$ for any choice of the potential. Plot (b) allows us to understand the behavior of $r$ as $\alpha$ increases. When $\alpha$ is around its minimum value $\alpha_{min}=\frac{k-4}{96\pi^2 k} \frac{N_e}{A_s} $ (which can be calculated by fixing $A_s$ in \eqref{eq:As:large:zeta:lnR} to its observed value), the scalar-to-tensor ratio is suppressed and increases very fast with $\alpha$ up to a maximum which depends on the choice of the potential. After reaching the maximum, $r$ is suppressed again, and for $\alpha \gtrsim 10^9$, all potentials predict the same $r$.  From (c), we see that all potentials exhibit similar behavior in $n_s$. All potentials predict $n_s = 0.983$ around $\alpha_{min}$, then it decreases up to a minimum value around $\alpha \sim 10^8$ and finally increases again towards the asymptotic value \eqref{eq:ns:R2lnR:small:zeta}. Notice that the asymptotic value for $n_s$ predicted by $V_e$ is $n_s = 1 - \frac{2}{N_e}$, which corresponds to the $k\rightarrow \infty$ limit in \eqref{eq:ns:R2lnR:small:zeta}. Finally, from (d), we have the relation between the two model parameters $\lambda,\alpha$ obtained by fixing $A_s$ to the observed value. We notice that for all potentials, increasing $\alpha$ implies increasing $\lambda$. We also notice that for all potentials, $\lambda$ drops to zero very fast as we approach $\alpha_{min}$.

Finally, notice that the $V_4$ potential is represented in (a) by the (blue) dot only, with coordinates $r\ll 1$ (b), $n_s = 0.95$ (c), since slow-roll inflation only happens for $\alpha \gtrsim 10^{13}$ when we already are in the strong coupling regime. We also see in (d) that the coupling $\lambda$ is fixed to $\lambda \sim 1.7\cdot 10^{-3}$ for all such $\alpha$ values.

\subsection{Back to quadratic gravity} \label{sec:back:to:R2}
\begin{figure}[t]
    \centering
\subfloat[]{\includegraphics[width=0.45\textwidth]{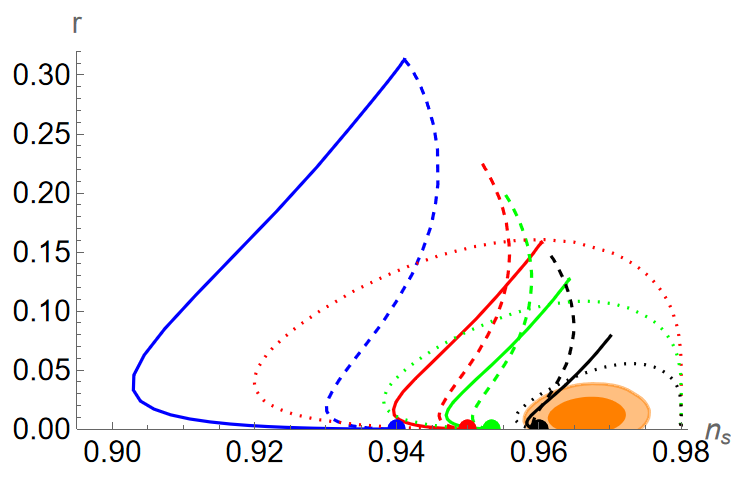}}
    \qquad
   \subfloat[]{\includegraphics[width=0.45\textwidth]{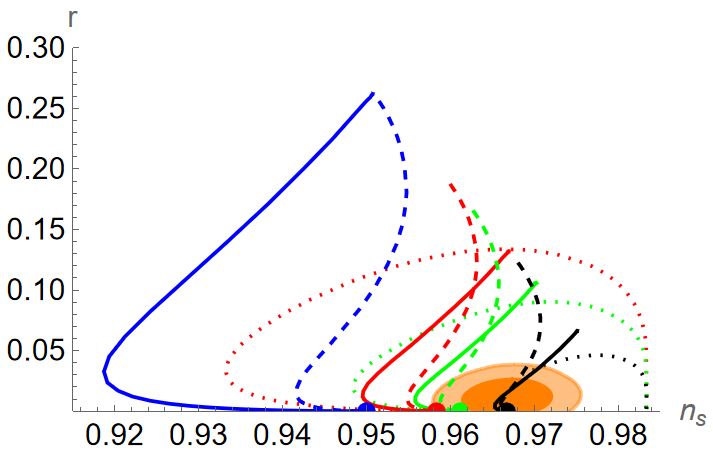}}%
    \caption{
    The observables $r$ vs. $n_s$ for $F(R) = R + \alpha R^3$ (thick), $F(R) = \frac{1}{\alpha}e^{\alpha R}$ (dashed), $F(R) = R + \alpha R^2 \ln(\alpha R)$ (dotted), and $V_k$ for $k=4$ (blue), $k=6$ (red), $k=8$ (green) and $V_e$ (black) for $N_e=50$ (left) and $N_e=60$ (right).  In the same color code, we show respectively the limit values for $\alpha \rightarrow \infty$, represented by bullets. The scalar amplitude $A_s$ is fixed to its observed value \cite{Planck2018:inflation}.
    The orange areas represent the 1,2$\sigma$ allowed regions coming  from  the latest combination of Planck, BICEP/Keck and BAO data \cite{BICEP:2021xfz}. It is clear from the plot that, for a given Jordan frame potential, all $\FTtwo$ predict the same $r, n_s$ in the quadratic $\alpha \rightarrow \infty$ limit.}
    \label{fig.r_vs_ns_all}
   \end{figure}

A summary of the results of the previous sections, including also the predictions for $N_e=50$, is presented in Fig.~\ref{fig.r_vs_ns_all}, where it can be seen that the higher number of $e$-folds $N_e=60$ is strongly favoured and that in the strong coupling limit, the slow-roll predictions of the $\FTtwo$ models are universal. In the following, we will prove that such a limit corresponds to a type of quadratic gravity.

 First of all, we note that since $G(\zeta_0)=0$, $\zeta=\zeta_0$  is the solution for the problem in the absence of matter, i.e. $V(\phi)=\partial_\mu \phi=0$ (cf. eq. \eqref{eq:EoMzetafull}). This case is well known \cite{Borowiec:1996kg} and leads to General Relativity plus the cosmological constant $U(\zeta_0)=\zeta_0/(4 F'(\zeta_0))$. In the  strong coupling limit $\alpha \to \infty$, $\zeta\rightarrow \zeta_0$ and inflation can be understood in general terms without specifying the full form of the $F(R)$. For this purpose, let us expand the $F(\zeta)$ function in Taylor series around $\zeta_0$ up to the quadratic order,
\bea
F_2(\zeta) &=& F(\zeta_0) + F'(\zeta_0)(\zeta-\zeta_0) + \frac{1}{2}F''(\zeta_0)(\zeta-\zeta_0)^2 \nn\\
&=& 2\Lambda - \omega \zeta + \alpha \zeta^2 \label{eq:FR2} \, ,
\eea
where 
\bea \label{eq:ident1}
\Lambda &=& -\zeta_0 G'(\zeta_0) \, , \\
\omega &=& -4 G'(\zeta_0) \, , \\\label{eq:ident2}
2\alpha &=& F''(\zeta_0)\label{eq:ident3} \, ,
\eea
and we used \eqref{eq:G} and $G(\zeta_0)=0$.
Given that we deal only with $\FTtwo$'s and the properties of $G(\zeta)$ explained below \eqref{eq:EoMzeta:SR}, it is easy to prove that $\Lambda,\omega,\alpha>0$. Therefore eq. \eqref{eq:FR2} exhibits a negative sign for the Einstein--Hilbert term. Such an unfamiliar configuration could be a priori allowed as long as the theory preserves the constraints
 $F_2'(R)>0$ and $F_2''(R) \neq 0$. Unfortunately, it can be shown that $F_2'$ stays positive only under slow-roll\footnote{This can be seen by combining \eqref{eq:FR2} with \eqref{eq:EoMzetafull} and solving the corresponding EoM for $\zeta$. Requiring $F'_2(\zeta)>0$ and considering a homogeneous $\phi$, the solution gives the constraint $\dot\phi^2 < \omega/2\alpha$, which will be broken with enough kinetic energy.}, therefore \eqref{eq:FR2} can only be used as an effective description for slow-roll inflation in the strong coupling limit. Nevertheless, the parametrization in \eqref{eq:FR2} is convenient to understand the results of the previous subsections, therefore we proceed with the inflationary computations. Using \eqref{eq:FR2} with  \eqref{eq:EoMzeta:SR}, we see that $\zeta$ now has the following solution in the slow-roll limit:
\be \label{eq:zetaquad:SR}
\zeta (\phi) = \frac{4\Lambda - 4V(\phi)}{\omega} \, ,
\ee
and the limit $\zeta \to \zeta_0$ can be interpred as $V(\phi) \to 0$.

Using \eqref{eq:zetaquad:SR} with the monomial potential \eqref{eq:V:k}, the field redefinition and the Einstein frame potential become, respectively,
\bea
\qty(\frac{\partial \chi}{\partial \phi})^2 &=& \frac{\omega}{8\alpha \left( \Lambda+\lambda_k\phi^k \right)-\omega^2} \, , \nn\\ 
U(\phi) &=&  \frac{ \lambda_k\phi^k+\Lambda}{ 8\alpha ( \lambda_k\phi^k+\Lambda)-\omega^2 } \, .
\eea
Since we are operating in the $V(\phi) \to 0$ limit, we only keep the leading order terms at small $\phi$ and obtain the following Einstein frame potential as a function of the canonically normalized field $\chi$:
\be
U(\chi) \simeq U_0 \qty(1-\Bar\lambda \chi^k) \label{eq:Ulimit:R2} \, ,
\ee
where
\bea
\label{eq:U0:R2}
U_0 &=&  \frac{1}{ 8\alpha} \frac{\Lambda}{\Lambda-\frac{\omega^2}{8\alpha}} \, , \\
\Bar{\lambda} &=& \frac{\lambda_k \, \omega  }{\Lambda} \left(\frac{8 \alpha  \Lambda }{\omega }-\omega
   \right)^{\frac{k}{2}-1} \, . \label{eq:lambda:bar}
\eea
Therefore, the potential \eqref{eq:Ulimit:R2} is nothing but the well-known $k$-hilltop potential (e.g. \cite{Boubekeur:2005zm,Tzirakis:2007bf,Lillepalu:2022knx} and refs. therein), which predicts
\bea
&&r \simeq \frac{2 \, U_0}{3 \pi^2 A_s} \simeq \frac{1}{12 \pi^2 A_s \alpha}  \ll 1 \, , \label{eq:r:R2:k}  \\%
&&n_s =1-\frac{k-1}{k-2}\frac{2}{N_e} \label{eq:ns:R2:k} \, ,
\eea
for $k \geq 3$ when $\bar\lambda \gg 1$. The last condition is equivalent to $\alpha \gg 1$. This explains why the $r$ and $n_s$ predictions are universal for any $\FTtwo$ in the strong coupling limit (cf. eqs. \eqref{eq:r:Rn:small:zeta}, \eqref{eq:ns:Rn:small:zeta}, \eqref{eq:r:eR:small:zeta}, \eqref{eq:ns:eR:small:zeta}, \eqref{eq:r:R2lnR:small:zeta} and \eqref{eq:ns:R2lnR:small:zeta}).

\section{Beyond slow-roll} \label{sec:beyondSR}
To better understand the global dynamics of our models, it is interesting to study their evolution numerically without the slow-roll approximation. The full Einstein frame EoMs read \cite{Dioguardi:2021fmr}
\begin{gather}
    \label{eq:phi_eom}
    \ddot{\phi} + 3H\dot{\phi} + \frac{V'(\phi)}{F'(\zeta)} = \frac{\dot{\phi}\dot{\zeta}F''(\zeta)}{F'(\zeta)}  \, , \\
    \label{eq:H_eom}
    3H^2 = \frac{1}{2}\frac{\dot{\phi}^2}{F'(\zeta)} + U(\phi,\zeta) \, , \\
    \label{eq:zeta_eom}
    -\frac{1}{2}\dot{\phi}^2 F'(\zeta) + 2V(\phi) - 2G(\zeta) = 0 \, . 
\end{gather}

The first slow-roll parameter can be written as
\begin{equation}
    \epsilon_H \equiv -\frac{\dot{H}}{H^2} = \frac{12V(\phi) - 6F(\zeta) + 3\zeta F'(\zeta)}{6V(\phi) - 3F(\zeta) + 2\zeta F'(\zeta)} \, .
\end{equation}
Since solving the constraint equation \eqref{eq:zeta_eom} is hard, we replace it with an equation for the time derivative of $\zeta$ \cite{Dioguardi:2021fmr},
\begin{equation} \label{eq:zeta_deriv}
    \dot{\zeta} = \frac{3H\dot{\phi}^2 F'(\zeta)  + 3 V'(\phi)\dot{\phi}}{2G'(\zeta) + \frac{3}{2}\dot{\phi}^2 F''(\zeta)} \, .
\end{equation}
As long as $\zeta$ solves \eqref{eq:zeta_eom} initially, equation~\eqref{eq:zeta_deriv} guarantees that \eqref{eq:zeta_eom} holds at all times.

\begin{figure}[t]
    \centering
    \includegraphics{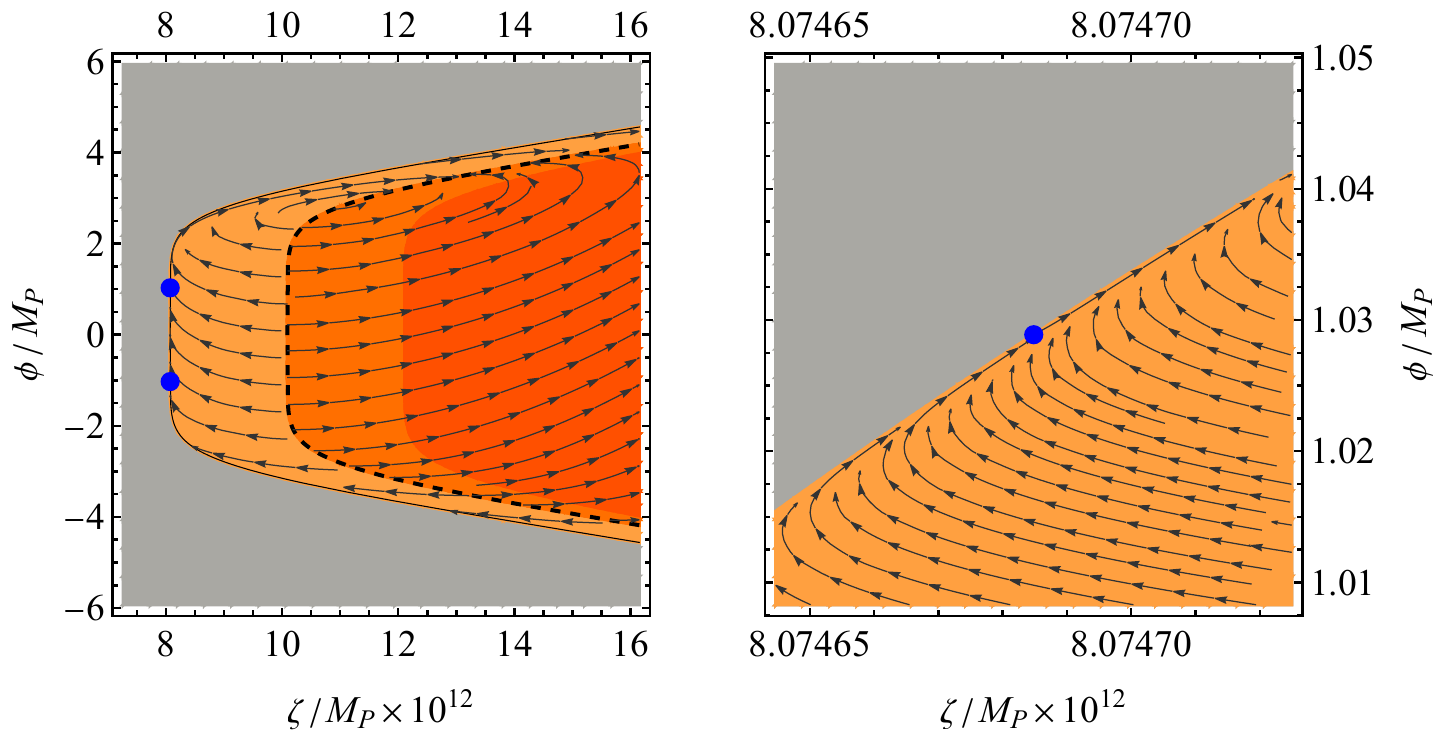}
    \caption{Flow of trajectories in the $\frac{1}{\alpha} \exp(\alpha R)$ model, globally (left) and zoomed into the slow-roll region (right). The arrows show time evolution in the $(\zeta,\phi)$ plane for the $\dot{\phi} > 0$, $H>0$ branch. The gray region is excluded by the constraint \eqref{eq:zeta_eom}, and slow-roll inflation takes place along its boundary, with the blue points indicating the CMB scale. The dashed line corresponds to the pole of $\dot{\zeta}$, with regions of inflation on either side. Inflation is only broken in the innermost dark orange region.
 }
    \label{fig:flow_exp}
\end{figure}

Figure~\ref{fig:flow_exp} presents the flow of trajectories in $(\zeta,\phi)$ space following eqs. \eqref{eq:phi_eom}, \eqref{eq:H_eom}, and \eqref{eq:zeta_deriv} for the exponential $F(R)$ described in Section \ref{sec:exp} and $V(\phi)=V_k (\phi)$ given in \eqref{eq:V:k} with $k=8$ at the benchmark point $\lambda=0.047$, $\alpha \simeq 2.48 \times 10^{11}$. This point realizes the quadratic limit for $F(R)$ with $n_s \simeq 0.961$,  $r \simeq 8.83 \times 10^{-6}$ at $N_e=60$. The flow plot is similar for all the models considered in this paper. The gray outer region is excluded because $V(\phi) < G(\zeta)$ and the constraint \eqref{eq:zeta_eom} cannot be satisfied there. At the edge of the region, $V(\phi) = G(\zeta)$ and $\dot{\phi}=0$, giving slow-roll inflation. Over time, trajectories deviate from this edge into the orange inner region, and $\dot{\phi}$ starts to grow. Unfortunately, in this limit, the extra kinetic terms complicate the time evolution, so there is no guarantee for a graceful exit from inflation. Let us study this analytically.

Inflation ends when $\epsilon_H>1$. Typically, $\epsilon_H$ reaches its maximum when $V(\phi)=0$ and $\dot\phi \to \infty$. This is true for the $F(R)=R+\alpha R^n$ model, and eq. \eqref{eq:zeta_eom} in this limit gives
\be
 \zeta \to \frac{n}{n-2} \dot\phi^2 \label{eq:zeta:Rn:beyond:SR} \, ,
\ee
and therefore 
\be
 \epsilon_{H \, \text{max}} = \frac{3 (n-2)}{2 n-3} \label{eq:epsH:Rn:beyond:SR} \, .
\ee
In the domain $n>2$, $\epsilon_H > 1$ can be realized if $n>3$. This is also true for $F(R)=\frac{1}{\alpha} e^{\alpha R}$, for which a similar analysis yields $\epsilon_{H \, \text{max}} = 3/2$, corresponding to the $n \to \infty$ limit of \eqref{eq:epsH:Rn:beyond:SR}. For $2 < n \leq 3$, we have $\epsilon_H < 1$ everywhere, and inflation never ends. The same is true for the $F(R)= R + \alpha R^2 \ln (\alpha R)$ model\footnote{Using a configuration with two free parameters, $F(R)= R + \alpha R^2 \ln (\beta R)$ solves this issue, but inflation would still end only in the region where slow-roll does not take place. See the next paragraph.}. These models are thus ruled out.

Unfortunately, even in the exponential and $n>3$ polynomial models another issue arises before the end of inflation is reached: there is a pole in $\dot\zeta$, corresponding to a pole in $\ddot\phi$, see eqs. \eqref{eq:zeta_deriv} and \eqref{eq:phi_eom}. This happens because in the denominator of \eqref{eq:zeta_deriv}, the first term is always negative ($G'<0$ for $\zeta>\zeta_0$), while the second term is either null or positive, and at a certain point, the terms cancel each other. When this pole is reached, the system breaks down. In all of our example cases, this happens while the system is still inflating, making it impossible to exit inflation gracefully.

To demonstrate this for the exponential case, we can solve the position of the pole from the denominator of \eqref{eq:zeta_deriv} as
\be
  \dot\phi^2 = \frac{ 1  }{3  } \left( \zeta-  \frac{1 }{ \alpha } \right) \label{eq:kin:pole} \, .
\ee
Using \eqref{eq:zeta_eom} and \eqref{eq:kin:pole}, we can write $\epsilon_H$ at the pole as a function of $\zeta$, obtaining
\be
 \epsilon_H = \frac{2-2 \alpha  \zeta }{1-2 \alpha  \zeta } \, ,
\ee
which is never bigger than one for $\zeta>\zeta_0=\frac{2}{\alpha}$. Since $\epsilon_H$ is a growing function of $\zeta$, it is smaller than one in the whole low-$\zeta$ slow-roll region. Inflation only ends in the large-$\zeta$ region, which can't be reached without hitting the pole\footnote{Slow-roll inflation typically ends before the pole is hit, with the second slow-roll parameter growing large. For our benchmark CMB point, the pole is hit 62 e-folds after the CMB exit, with $\epsilon_H=0.82$.}.

Therefore, in its current form, our scenario can only work as an effective theory for the slow-roll regime. New physics is required to either remove the pole from the equation of motion of $\zeta$ or to gracefully exit inflation before the insurgence of the pole.

On the other hand, the region beyond the pole has interesting features. If the initial $\zeta$ (i.e. the initial kinetic energy) is small enough, the system inflates (without slow-roll), then exits inflation, re-enters another non-slow-roll inflationary phase at a later time, and remains there. Such a configuration seems promising for a joint explanation of the Universe's early and late accelerated expansion. Such quintessential inflation models have been studied in the context of Palatini $R^2$ gravity in \cite{Dimopoulos:2022tvn, Dimopoulos:2022rdp}. However, further studies are needed and postponed to a separate dedicated work.

\section{Beyond $F(R)$: $F(R,X)$}\label{sec:beyond_F(R)}

It is clear from the previous section that $F(R)$ theories do not provide a graceful exit from inflation. The source of the problem is the $F'$ contribution in \eqref{eq:dchidphi}, which makes $\zeta$ a function also of the derivatives of $\phi$ with an ultimate effect of creating the pole in eq. \eqref{eq:zeta_deriv}. There is, however, a simple way to fix the issue, which relies on generalizing the $F(R)$ function to $F(R,X)$, with $X = -g^{\mu\nu}\partial_\mu\phi\partial_\nu\phi$ the inflaton kinetic term \cite{Dioguardi:2023}. To see how this works, consider the action 
\be
S_J = \int d^4x \sqrt{-g^J}\qty[\frac{1}{2}F(R_{\gamma X}) - V(\phi)] \, ,
\ee
where $R_{\gamma X} = R + \gamma X$ and $\gamma$ is a free parameter\footnote{In general, for the sole purpose of having the canonical normalization of the scalar field independent of $\zeta$, $\gamma$ could also be a function of $\phi$. However, for the purposes of our current investigation it is enough to consider $\gamma$ a constant free parameter. We stress that the value $\gamma=0$ is not allowed, because it would completely remove the kinetic term of $\phi$ from the action, making $\phi$ another auxiliary field and not anymore the inflaton.}.  We can rewrite the action in the following way by introducing an auxiliary field $\zeta$:
\be
S_J = \int d^4x\sqrt{-g^J}\qty[\frac{1}{2}F(\zeta) + \frac{1}{2}F'(\zeta) (R - \gamma g^{\mu\nu}_J \partial_\mu \phi\partial_\nu \phi -\zeta) - V(\phi)] \, ,
\label{eq:action:zeta:J:FRX}
\ee
which straightforwardly yields
\be 
  S_J = \int \dd^4 x \sqrt{-g}_J \left[\frac{1}{2} \left[F(\zeta)+F'(\zeta) \left(R(\Gamma)-\zeta \right) \right] - \frac{1}{2}\gamma F'(\zeta) g_J^{\mu\nu} \partial_\mu \phi \partial_\nu \phi - V(\phi) \right] \, .
\label{eq:action:zeta:J:FRX2}
\ee
Note that action \eqref{eq:action:zeta:J:FRX2} differs from \eqref{eq:action:zeta:J} only for a $\gamma F'(\zeta)$ prefactor in front of the inflaton kinetic term.
Analogously, applying the Weyl rescaling \eqref{eq:Weyl} to the action \eqref{eq:action:zeta:J:FRX2}, we obtain the Einstein frame action
\be \label{eq:action:zeta:E:FRX}
S_E = \int d^4x \sqrt{-g^E} \qty[\frac{R}{2} - \frac{1}{2}g^{\mu\nu}_E \partial_\mu \chi \partial_\nu \chi - U(\zeta,\chi)] \, ,
\ee
where $U(\zeta,\chi)$ is still given by eq. \eqref{eq:Uchizeta}. Hence from \eqref{eq:action:zeta:J:FRX2} and \eqref{eq:action:zeta:E:FRX} we see that 
\be
\label{eq:dchidphi:RX}
 \frac{\pd \chi}{\pd \phi} = \sqrt{\gamma  \frac{F'(\zeta)}{F'(\zeta)}}=\sqrt{\gamma} \, ,
\ee
and the $F'$ prefactor of the inflaton kinetic term is canceled out by the $F'$ factor coming from the Weyl transformation, leaving the canonical normalization of the inflaton depending only on $\gamma$.
Therefore the action \eqref{eq:action:zeta:E:FRX} differs from \eqref{eq:action:zeta:E} only for the normalization of the inflaton kinetic term: now only a global rescaling is needed to make it canonically normalized.
As we will see later, this implies that there is no pole in $\dot\zeta$ making the evolution of the field smooth and allowing for a graceful exit from the inflation era. Moreover, the $\zeta$ EoM
\be \label{eq:G=V}
G(\zeta) = \frac{1}{4}\qty[2 F(\zeta) - \zeta F'(\zeta)] = V(\phi)
\ee
is now exact and not anymore an approximation only valid under slow-roll (cf. \eqref{eq:EoMzetafull}). 

\subsection{Slow-roll predictions} 
The inflationary computations will still proceed following the method introduced in \cite{Dioguardi:2021fmr}. However, in the $F(R_{\gamma X})$ case, there is one more constraint to satisfy. The slow-roll parameter $\epsilon(\zeta)$ reaches a local maximum since $U(\phi)$ differs from zero at its absolute minimum. We need to enforce $\epsilon > 1$ at the local maximum in order to have a graceful exit. This can be accomplished by increasing $\alpha$, establishing an $\alpha_{min}$ such that the end of slow-roll inflation can be achieved. The value of $\alpha_{min}$ depends both on the choice of $F(\zeta)$ and $V(\phi)$ and has to be evaluated numerically. However, as a general feature, $\alpha_{min}$ lies in the strong coupling limit. Viable inflation then only takes place in the quadratic $F$ limit \cite{Dioguardi:2023}.

We proceed analogously to subection \ref{sec:back:to:R2}, using the $F_{2}(\zeta)$ quadratic limit \eqref{eq:FR2}. The only difference is that $\zeta$ is now given by the combination $\zeta = R+\gamma X$ in the Jordan frame. This class of theories has been preliminarily studied in \cite{Dioguardi:2023}  for the potentials given in eqs. \eqref{eq:V:k} and \eqref{eq:V:exp}, and it predicts again the same $k$-hilltop results \eqref{eq:r:R2:k} and \eqref{eq:ns:R2:k}, independently of $\gamma$. In such a limit, the dependence on $\gamma$ is only visible in the normalization of the scalar potential $V(\phi)$, affecting the value of $A_s$. This can be shown by following the procedure of subsection \ref{sec:back:to:R2}, obtaining
\be
U(\chi) \simeq U_0 \qty(1-\lambda_\gamma \chi^k)  \, ,
\label{eq:Ulimit:R2:2}
\ee
where $U_0$ is given in \eqref{eq:U0:R2} and
\be
\lambda_\gamma = \frac{\lambda_k\omega}{\Lambda}\frac{\omega \gamma^{-k/2}}{8\alpha\Lambda-\omega^2} \, . 
\ee
Setting $\gamma = 1/F'(\zeta_0)$ (compare eq. \eqref{eq:dchidphi:RX} to the $\zeta \to \zeta_0$ limit of eq. \eqref{eq:dchidphi}), this coincides with the previous result \eqref{eq:lambda:bar}, and the potentials \eqref{eq:Ulimit:R2:2} and \eqref{eq:Ulimit:R2} become equal. 

 To conclude, we point out that the $\alpha_{min}$ value that allows for the end of slow-roll inflation is such that $r\lesssim 10^{-5}$ \cite{Dioguardi:2023}, implying that the tensor-to-scalar ratio would not be measurable even by a futuristic high-resolution satellite mission such as PICO \cite{NASAPICO:2019thw}. Therefore, in the strong coupling limit, we cannot see any difference between the predictions of $F_{>2}(R)$ and $F_{>2}(R_{\gamma X})$, but the latter fixes the inconvenience of non-minimal kinetic terms and the lack of a graceful exit, as we will demonstrate in the next subsection. 

\subsection{Beyond slow-roll}
In this subsection, we briefly discuss the dynamics beyond slow roll in the $F(R_{\gamma X})$ setup.
In this case, the full Einstein frame EoMs in terms of the Jordan frame scalar field $\phi$ and the auxiliary field $\zeta$ read:
\bea
&&\ddot \phi + 3 H \dot\phi + \frac{V'(\phi)}{\gamma F'(\zeta)^2}=0 \, \label{eq:phi_eom:RX},\\%
&&3H^2 =\frac{1}{2}\gamma \dot\phi^2 + U(\phi,\zeta) \, ,\label{eq:H_eom:RX} \\%
&&V(\phi) = G(\zeta) \label{eq:zeta_eom:RX} \, . 
\eea
We stress the differences between the current set of EoMs and the ones previously obtained in section \ref{sec:beyondSR}, apart from the obvious contribution from $\gamma$. With respect to eq. \eqref{eq:phi_eom}, a $\dot\zeta$ term is missing from the right-hand side of \eqref{eq:phi_eom:RX}, because the $\phi$ kinetic term is independent of $\zeta$ (see eqs. \eqref{eq:action:zeta:E:FRX} and \eqref{eq:dchidphi:RX}). For the same reason, a $\dot\phi$ term is missing from the left-hand side of \eqref{eq:zeta_eom:RX}, with respect to the EoM \eqref{eq:zeta_eom}. The disappearance of these two terms makes the evolution of the canonical field $\chi$ smooth. This can be seen easily, by rewriting the EoMs \eqref{eq:phi_eom:RX}, \eqref{eq:H_eom:RX} and \eqref{eq:zeta_eom:RX}  in terms of the canonical field $\chi$. Using  the field redefinition \eqref{eq:dchidphi:RX} we straightforwardly obtain
\bea
&&\ddot \chi +3 H \dot\chi + U'(\chi)=0 \, , \label{eq:phi_eom:RX:chi}\\
&&3H^2 =\frac{1}{2} \dot\chi^2 + U(\chi) \, , \label{eq:H_eom:RX:chi} \\
&&V(\chi) = G(\zeta) \label{eq:zeta_eom:RX:chi} \, ,
\eea 
where $U(\chi) = U(\chi,\zeta(\chi))$ with $U(\chi,\zeta)$ given in eq. \eqref{eq:Uchizeta} and $\zeta(\chi)$  the solution\footnote{The computational method introduced in \cite{Dioguardi:2021fmr} relies on the fact that such a solution exists, but we cannot explicitly write it down. However, we are still allowed to write it formally as we do here.} of the auxiliary field EoM in eq. \eqref{eq:zeta_eom:RX:chi}.
One can immediately notice that eqs. \eqref{eq:phi_eom:RX:chi} and \eqref{eq:H_eom:RX:chi} are the standard EoMs for the inflaton and the Hubble parameter, while eq. \eqref{eq:zeta_eom:RX:chi} leads to the simple expression
\be
\dot \zeta = \frac{V'(\chi)\dot\chi}{G'(\zeta)} \, .
\ee
In the region of interest, $\zeta>\zeta_0$, there is no pole\footnote{Mathematically, a pole is present when $G'=0$. This corresponds to the local maximum of $G$ (see Fig. \ref{fig:GvsV}). The appearance of the pole in this case represents the existence of two possible solutions satisfying the EoM for $\zeta$. However, since we restrict ourselves to working in the $\zeta>\zeta_0$ region, such a danger is avoided.} in $\dot{\zeta}$. Evolution is smooth even outside of slow roll. To see that a graceful exit is guaranteed (provided that $\alpha>\alpha_\text{min}$ ensuring first the end of slow-roll), we note that
\be
\epsilon_H = -\frac{\dot H}{H^2} = \frac{3\dot\chi^2/2}{\dot\chi^2/2 + U(\chi)} \, ,
\ee
hence $\epsilon_H > 1$ for a large enough kinetic term, independently of the model. 

We stress once again that the $F(R_{\gamma X})$ setup does not change the form of the potential $U(\chi,\zeta)$ (cf. eqs. \eqref{eq:action:zeta:E},  \eqref{eq:EoMzeta:SR}, \eqref{eq:action:zeta:E:FRX} and \eqref{eq:G=V}), hence the inflationary computations under the slow-roll regime are the same and yield the same observables. However, from \eqref{eq:action:zeta:E:FRX}, it is clear that the higher order kinetic terms that would spoil the dynamics around the end of slow-roll here are not present. This allows for a graceful exit from the inflationary era without affecting the form of $U(\chi,\zeta)$.

\section{Conclusions} \label{sec:Conclusions}
We studied single-field slow-roll inflation embedded in Palatini $F(R)$ gravity. Surprisingly, for $F(R)$ that grow faster than $R^2$, the consistency of the theory requires the Jordan frame inflaton potential to be unbounded from below. Even more surprisingly, the corresponding Einstein frame potential is bounded from below, positive definite, and has a plateau at small inflaton field values and a tail approaching the horizontal axis at big inflaton values. We studied the phenomenology at small field values, considering three test scenarios for this kind of gravity, $F(R)=R+\alpha R^n$, $F(R)=\frac{1}{\alpha} \exp(\alpha R)$ and $F(R)=R+\alpha R^2 \ln (\alpha R)$, and computed the inflationary predictions with $V_k(\phi)$, a negative monomial potential, and $V_e(\phi)$, a negative exponential potential, for $N_e =50, 60$ $e$-folds.  We found that the steeper the potential, the better the agreement with the experimental constraints \cite{BICEP:2021xfz}, with the negative exponential potential the most favoured of all the cases considered. Finally, $N_e=60$ is strongly favoured over $N_e=50$.  Moreover, we proved that for all the Palatini $F(R)$ that diverge faster than $R^2$, there exists a universal strong coupling limit configuration corresponding to a quadratic $F(R)$ with the $\emph{wrong}$ sign for the linear term and a cosmological constant in the Jordan frame. In such a limit, the tensor-to-scalar ratio $r$ does not depend on the original inflaton potential $V(\phi)$, but only on the effective coupling of the quadratic term in $R$. On the other hand, the scalar spectral index $n_s$ depends on the original potential $V(\phi)$. 

Unfortunately, the study of the dynamics beyond the slow-roll regime shows the insurgence of a pole in the equation for $\dot{\zeta}$ before the end of inflation. To solve this issue, we extended the model to the class of $F(R,X)$ theories with $X$ the inflaton kinetic term. The strong coupling predictions for this class of theories for slow-roll inflation are the same as for the $F(R)$ ones. However, with $F(R,X)$, we can eliminate the pole in $\dot\zeta$ and guarantee a graceful exit in the strong coupling limit. We also get rid of the cumbersome higher-order kinetic terms present in the Einstein frame in the $F(R)$ case.

Before concluding, we stress that essentially model independently, any $F(R,X)_{>2}$ provides an inflaton potential with a tail approaching the horizontal axis. This could provide a natural explanation for the smallness of the cosmological constant \cite{Planck:cosmo}. Further studies are needed in this direction and will follow in a future work.

\acknowledgments


This work was supported by the Estonian Research Council grants PRG1055,  RVTT3, RVTT7 and the CoE program TK202 ``Fundamental Universe'’.

\appendix

\bibliographystyle{JHEP}
\bibliography{references}

\end{document}